\documentclass[epj]{svjour}
\usepackage{graphics}
\begin{document}
\title{Electrodisintegration of $^3$He below and above deuteron breakup 
threshold}
\author{L.E. Marcucci\inst{1,2} 
\and M. Viviani\inst{2,1}
\and R. Schiavilla\inst{3,4}
\and A.Kievsky\inst{2,1}
\and S. Rosati\inst{1,2}
}                     
\institute{Department of Physics, University of Pisa, I-56100 Pisa, Italy 
\and INFN, Sezione di Pisa, I-56100 Pisa, Italy
\and Department of Physics, Old Dominion University, Norfolk, 
Virginia 23529, USA
\and Jefferson Lab, Newport News, Virginia 23606, USA}
\date{Received: date / Revised version: date}
%
\abstract{
Recent advances in the study of electrodisintegration of $^3$He 
are presented and discussed. The pair-correlated hyperspherical harmonics
method is used to calculate the initial and final state wave functions, 
with a realistic Hamiltonian consisting of the Argonne $v_{18}$ 
two-nucleon and Urbana IX three-nucleon interactions. The model 
for the nuclear current and charge operators 
retains one- and many-body  contributions.
Particular attention is made in the construction of the two-body current 
operators arising from the momentum-dependent part of the two-nucleon 
interaction. Three-body current operators are also included 
so that the full current operator is strictly conserved. 
The present model for the nuclear 
current operator is tested comparing theoretical predictions and 
experimental data of $pd$ radiative capture cross section and 
spin observables.
\PACS{
      {21.45.+v}{Few-body systems}   \and
      {25.10.+s}{Nuclear reactions involving few-nucleon systems} \and 
      {25.30.Dh}{Inelastic electron scattering to specific states} \and
      {25.30.Fj}{Inelastic electron scattering to continuum} \and
      {25.40.Lw}{Radiative capture}
     } 
} 
\maketitle
\section{Introduction}
\label{intro}

The theoretical study of the electromagnetic structure of few-body 
nuclei requires the knowledge of the nuclear wave functions and
electromagnetic transition operators. In the case of processes 
involving two and/or three nucleons, it is possible to 
obtain very accurate bound- and scattering-state wave functions from 
realistic Hamiltonian models. Therefore, the different models for the 
nuclear electromagnetic current operator can be tested with the large 
variety of electromagnetic observables involving $A$=2 and 3 nuclei. 
In particular, we concentrate our attention on the $pd$ radiative 
capture and on the electrodisintegration of $^3$He below and above 
deuteron breakup threshold (DBT). 
These processes have been extensively studied 
by several research groups (see Ref.~\cite{Car98} for a review). 
Most recently, the $pd$ radiative capture and the $^3$He($e,e'$) 
reactions below DBT have been investigated by our group in 
Ref.~\cite{Viv00}. The pair-correlated hyperspherical harmonics (PHH)
method~\cite{Kieetal} has been used  
to calculate the $A$=3 bound- and scattering-state wave functions 
from a realistic Hamiltonian model consisting of the Argonne $v_{18}$ 
two-nucleon~\cite{Wir95} and Urbana IX three-nucleon~\cite{Pud95} 
interactions (AV18/UIX). The nuclear electromagnetic current operator 
included, in addition to the one-body convection and spin-magnetization 
terms, also two-body contributions. These two-body terms were constructed 
following the method of Ref.~\cite{Ris85}, with the goal of 
satisfying the current conservation relation (CCR) with the AV18. 
However, within 
this method, only the dominant 
two-body terms, constructed from the momentum-independent part of the AV18, 
satisfy the CCR with this part of the interaction~\cite{Sch89}. 
The two-body terms 
originated from the momentum-dependent part of the AV18 are not strictly 
conserved. 
In Ref.~\cite{Viv00}, further transverse contribution, associated with the 
$\rho\pi\gamma$ and $\omega\pi\gamma$ transition mechanisms and with the 
excitation of intermediate $\Delta$ resonances, were included. 

The main conclusions of Ref.~\cite{Viv00} can be summarized as follows: 
the two-body contributions to the electromagnetic 
charge and current operators are
crucial to achieve an overall good agreement between theory and 
experiment for all the observables under investigation. 
However, some discrepancies between theory and experiment have been found, 
in particular for the deuteron tensor polarization 
observables $T_{20}$ and $T_{21}$ of $pd$ radiative capture 
at center-of-mass energy 
$E_{c.m.}=2$ MeV, and, in Ref.~\cite{Mar03}, also 
at $E_{c.m.}=3.33$ MeV.
In the analysis of Ref.~\cite{Viv00} it has been suggested that 
these discrepancies might be due to the fact that the 
electromagnetic current operator satisfies only approximately the CCR 
with the nuclear Hamiltonian. 
In fact, when the $T_{20}$ and $T_{21}$ observables are 
calculated in the 
long-wavelength approximation (LWA), applying the Siegert theorem, 
a quite good agreement with the data is obtained. 

In this work we present a new model for the nuclear 
current operator which {\it exactly} satisfies the CCR with 
the AV18/UIX Hamiltonian model. The model is then tested in the study of 
the $pd$ radiative capture at the two previously considered 
values of $E_{c.m.}$, 2 and 3.33 MeV, and in the 
study of inclusive and exclusive electrodisintegration of $^3$He, 
both below and above DBT. 
The model for the nuclear electromagnetic current and charge operators is 
summarized in the following section. A detailed review 
will be given elsewhere~\cite{Marip}. Finally, the theoretical results 
are compared with the experimental data in Sec.~\ref{sec:res}. 

\section{The nuclear electromagnetic charge and current operators}
\label{sec:em}

The nuclear electromagnetic charge $\rho({\bf q})$ and current 
${\bf j}({\bf q})$ operators can be written as sums of one- and 
many-body terms that operate on the nucleon degrees of freedom. 
The one-body operators $\rho_i({\bf q})$ and 
${\bf j}_i({\bf q})$ for particle $i$ 
are derived from the non-relativistic reduction
of the covariant single-nucleon current, by expanding
in powers of $1/m$, $m$ being the nucleon mass~\cite{Car98}.
The model commonly used~\cite{Sch90} for the 
two-body charge operators includes the $\pi$-, $\rho$-, and
$\omega$-meson exchange terms with both isoscalar and isovector
components, as well as the (isoscalar) $\rho \pi \gamma$ and (isovector)
$\omega \pi \gamma$ charge transition couplings. 
At moderate values of momentum transfer ($q \!< \! 5$ fm$^{-1}$), 
the $\pi$-meson exchange charge operator has been found to give 
the dominant two-body contribution~\cite{Mar98}.  

The electromagnetic current operator must satisfy the CCR, written as
\begin{equation}
  {\bf q}\cdot{\bf j}({\bf q})= [H,\rho({\bf q})]\ ,\label{eq:ccr}
\end{equation}
where the nuclear Hamiltonian $H$ is taken to consist 
of two- and three-body interactions, denoted as
$v_{ij}$ and $V_{ijk}$ respectively.
To lowest order in $1/m$, Eq.~(\ref{eq:ccr}) separates into
\begin{eqnarray}
  {\bf q}\cdot{\bf j}_i({\bf q})&=& 
\biggl[{{\bf p}_i^2\over 2m},\rho_i({\bf q})\biggr]
    \ ,\label{eq:ccr1}\\
  {\bf q}\cdot{\bf j}_{ij}({\bf q})&=& [v_{ij},\rho_i({\bf q})+\rho_j({\bf q})]
 \ , \label{eq:ccr2}
\end{eqnarray}
and similarly for the three-body current ${\bf j}_{ijk}({\bf q})$.
We have neglected the two-body terms in $\rho({\bf q})$, which 
are of order $1/m^2$.  The one-body current
is easily shown to satisfy Eq.~(\ref{eq:ccr1}).
To construct the two-body current, 
it is useful to adopt the classification scheme of Ref.~\cite{Ris89},
and separate the current ${\bf j}_{ij}({\bf q})$ into model-independent (MI)
and model-dependent (MD) parts. 
The MI two-body current has a longitudinal
component and is constructed so as to satisfy the CCR of Eq.~(\ref{eq:ccr2}), 
while the MD two-body current
is purely transverse and therefore is
un-constrained by the CCR.  The latter is taken
to consist of the isoscalar $\rho\pi\gamma$
and isovector $\omega\pi\gamma$ transition currents, as well as the
isovector current associated with excitation of intermediate
$\Delta$ resonances as in Ref.~\cite{Viv00}.

The MI two-body currents arising from the momentum-independent terms 
of the AV18 two-nucleon interaction have been constructed 
following the standard procedure of Ref.~\cite{Ris85}, 
which will be hereafter quoted as meson-exchange (ME) scheme.
It can be 
shown that these two-body current operators satisfy {\it exactly} the CCR
with the first six operators of the AV18. 
The two-body currents arising from the spin-orbit components of 
the AV18 could be constructed using again ME
mechanisms~\cite{Car90}, 
but the resulting currents turn out to be not strictly conserved. 
The same can be said of those currents deriving 
from the quadratic momentum-dependent components of the AV18, 
if obtained, as in Ref.~\cite{Viv00}, 
by gauging only the momentum operators, 
but ignoring the implicit momentum dependence which comes through
the isospin exchange operator (see below).
Since our goal is to construct MI two-body currents 
which satisfy {\it exactly} the CCR of Eq.~(\ref{eq:ccr2}), 
the currents arising from the momentum-dependent 
terms of the AV18 interaction have been obtained following the 
procedure of Ref.~\cite{Sac48}, which will be quoted 
as minimal-substitution (MS) scheme.
The main idea of this procedure, 
reviewed and extended in Ref.~\cite{Marip}, is that 
the isospin operator ${\bf \tau}_i\cdot{\bf \tau}_j$, 
is formally 
equivalent to an implicit momentum dependence~\cite{Sac48}. In fact, 
${\bf \tau}_i\cdot{\bf \tau}_j$ can be expressed in terms of the 
space-exchange operator ($P_{ij}$) using the formula
\begin{equation}
  {\bf \tau}_i\cdot{\bf \tau}_j = -1-
  (1+{\bf \sigma}_i\cdot{\bf \sigma}_j) P_{ij} \ ,
  \label{eq:tt}
\end{equation}
valid when operating on antisymmetric wave functions.
The operator $P_{ij}$ is defined as 
$P_{ij}=   {\rm e}^{{\bf r}_{ji}\cdot{\bf\nabla}_i
  + {\bf r}_{ij}\cdot{\bf\nabla}_j}$, where 
the ${\bf\nabla}$-operators do not act on the 
vectors ${\bf r}_{ij}={\bf r}_i-{\bf r}_j=-{\bf r}_{ji}$ in the exponential.
In the presence of an electromagnetic
field, minimal substitution is performed both in the 
momentum dependent terms of the two-nucleon interaction and in the  
space-exchange operator $P_{ij}$ of Eq.~(\ref{eq:tt}), 
and the resulting current operators are then obtained with standard 
procedures~\cite{Marip,Sac48}. 
Explicit formulas can be found in Ref.~\cite{Marip}. 
Here we only quote the result for the currents arising 
from the momentum-independent part of the AV18 interaction ($v_{ij}^0$):
\begin{equation}
  {\bf j}_{ij}({\bf q}) ={\rm i}\,v_{ij}^0\bigg(
  \epsilon_i\int_{\gamma_{ij}}d{\bf s}\,{\rm e}^{{\rm i}{\bf q}\cdot{\bf s}}
 +\epsilon_j\int_{\gamma'_{ji}}d{\bf s}'\,
{\rm e}^{{\rm i}{\bf q}\cdot{\bf s}'}\bigg)
  \,(1+{\bf \tau}_i\cdot{\bf \tau}_j) \ ,
\label{eq:jq}
\end{equation}
where $\epsilon_i$ is the nucleon charge operator, 
and $d{\bf s}$ ($d{\bf s}'$) is 
the infinitesimal step on the generic path $\gamma_{ij}$ 
($\gamma'_{ji}$) that 
goes from position $i$ ($j$) to position $j$ ($i$).
Two observations are in order: (i) with a particular 
choice of the integration path it is possible to re-obtain the 
two-body current operators calculated, within the ME scheme, from 
the momentum-independent part of the AV18~\cite{Marip}; (ii) in the limit  
${\bf q}\rightarrow 0$, the current operator 
${\bf j}_{ij}({\bf q})$ of Eq.~(\ref{eq:jq}) becomes 
path-independent and unique. 
To simplify the calculation, 
the integration paths $\gamma_{ij}$ and $\gamma'_{ji}$ have been 
chosen to be linear. 

Both the ME and the MS schemes 
can be generalized to calculate the three-body current operators 
induced by the three-nucleon interaction (TNI) $V_{ijk}$. 
Here, these three-body currents have been constructed within the ME scheme 
to satisfy the CCR with the 
Urbana-IX TNI~\cite{Pud95}. However, the same procedure can be 
applied to other models of TNI's, such as 
the Tucson-Melbourne~\cite{Coo79} and Brazil~\cite{Rob} models.
Details of the calculation can be found in Ref.~\cite{Marip}.

In summary, the present model for the many-body current operator 
retains 
the two-body terms obtained within the ME scheme from the 
momentum-independent 
part of the AV18, those ones obtained within the MS 
scheme from the momentum-dependent part of the AV18, the MD
terms quoted above, and the three-body terms obtained 
within the ME scheme from the UIX. Thus, the full current operator satisfies 
{\it exactly} the CCR with the AV18/UIX nuclear Hamiltonian. 
In contrast, the model of Ref.~\cite{Viv00} 
retains only two-body currents, all of them obtained within the 
ME scheme.

\begin{figure}
\vspace*{0.8cm}
\resizebox{0.45\textwidth}{!}{%
\includegraphics{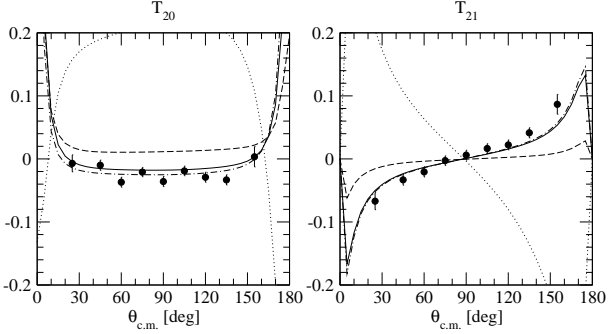}
}
\caption{Deuteron tensor polarization observables $T_{20}$ and $T_{21}$ 
for $pd$ radiative capture at $E_{c.m.}$= 2 MeV, 
obtained with the AV18 Hamiltonian models. 
See text for explanations of the different curves. 
The experimental data are from Ref.~\protect\cite{Smi99}.}
\label{fig:t20-t21}
\end{figure}

\begin{figure}
\vspace*{1cm}
\resizebox{0.38\textwidth}{!}{%
\includegraphics{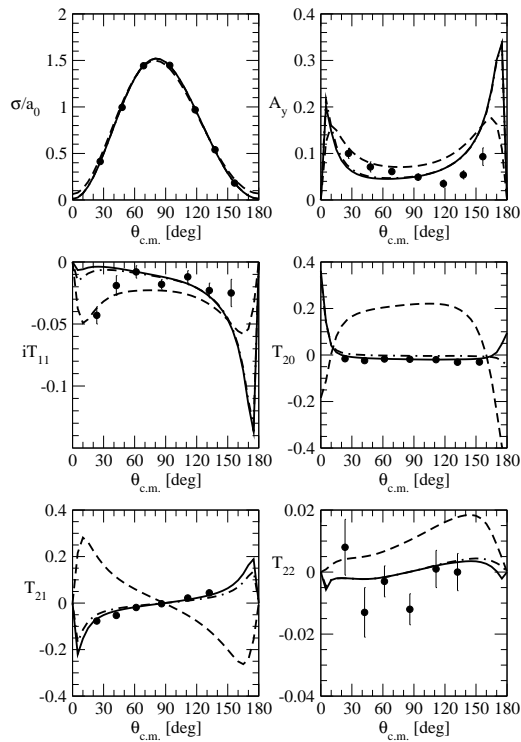}
}
\caption{Differential cross section, proton vector analyzing power, 
and the four deuteron tensor polarization observables 
for $pd$ radiative capture at $E_{c.m.}=$3.33 MeV, obtained with the 
AV18/UIX Hamiltonian model. 
See text for the explanations of the different curves.
The experimental data are from Ref.~\protect\cite{Goe92}.}
\label{fig:obs3.33}
\end{figure}

\section{Results}
\label{sec:res}

In the present section we compare the theoretical prediction and 
experimental data for two sets of $^3$He electrodisintegration observables: 
(i) the longitudinal and 
transverse response functions $R_L$ and $R_T$ for $^3$He($e,e'$) 
at momentum transfer values   
$q$=0.88, 1.64 and 2.47 fm$^{-1}$ and excitation energies ($E_x$) 
from two-body 
threshold up to 20 MeV; (ii) the differential cross section of the 
$^3$He($e,e'd$)$p$ reaction  as function of the missing momentum in 
($q,\omega$)-constant kinematics, at beam energies of 370 and 576 MeV 
and $q$ values of 412, 504 and 604 MeV/c. Data are respectively from 
Refs.~\cite{Ret94} and~\cite{Spa02}.

However, preliminarly, we report some results for the $pd$ radiative 
capture reaction, to show that the new model for the many-body nuclear 
current operators resolves some of the discrepancies 
between theory and experiment of Ref.~\cite{Viv00}.
The $T_{20}$ and $T_{21}$ observables at $E_{c.m.}$=2 MeV using the 
AV18 two-nucleon interaction are shown in Fig.~\ref{fig:t20-t21}. The dotted 
curves are obtained including only the one-body current 
contributions, the dashed curves are obtained with the model 
for the nuclear current operator of Ref.~\cite{Viv00}, and 
the dotted-dashed curves are obtained 
in the long-wavelength-approximation (LWA), applying the Siegert theorem. 
Finally, the solid curves are obtained  
including the one- and two-body contributions described in 
Sec.~\ref{sec:em}, necessary to satisfy the CCR with the AV18 
nuclear Hamiltonian. Data are from Ref.~\cite{Smi99}.
By inspection of Fig.~\ref{fig:t20-t21}, 
we observe that there is a good agreement between the experimental 
data, the LWA and the present  
``full'' results, while the results of 
Ref.~\cite{Viv00} are in disagreement, as expected, with both the 
LWA and experimental results. The excellent agreement 
between the LWA and the ``full'' results is a consequence 
of the fact that the ``full'' nuclear electromagnetic current 
operator satisfies the CCR with the AV18 nuclear Hamiltonian.

As an example of the degree of agreement which has been reached 
between the present calculation and experimental data for the 
$pd$ radiative capture, we show
in Fig.~\ref{fig:obs3.33} the theoretical predictions obtained 
with the AV18/UIX Hamiltonian model at $E_{c.m.}$=3.33 MeV. Data 
are from Ref.~\cite{Goe92}.
The dashed, dotted-dashed and solid curves correspond
to the calculation with one-body only, with one- and 
two-body, and with one-, two- and three-body 
currents.
An overall nice description has been reached for 
all the observables, with the only exception of the i$T_{11}$ 
deuteron polarization observable at small angles. Also, some 
small three-body currents effects are noticeable, especially in 
the $T_{20}$ and $T_{21}$ deuteron tensor observables, 
which is an indication of the fact that if a Hamiltonian 
model with two- and three-nucleon interactions is used, then the 
model for the nuclear current operator should include the corresponding 
two- and three-body contributions.
\begin{figure}
\resizebox{0.48\textwidth}{!}{%
\includegraphics{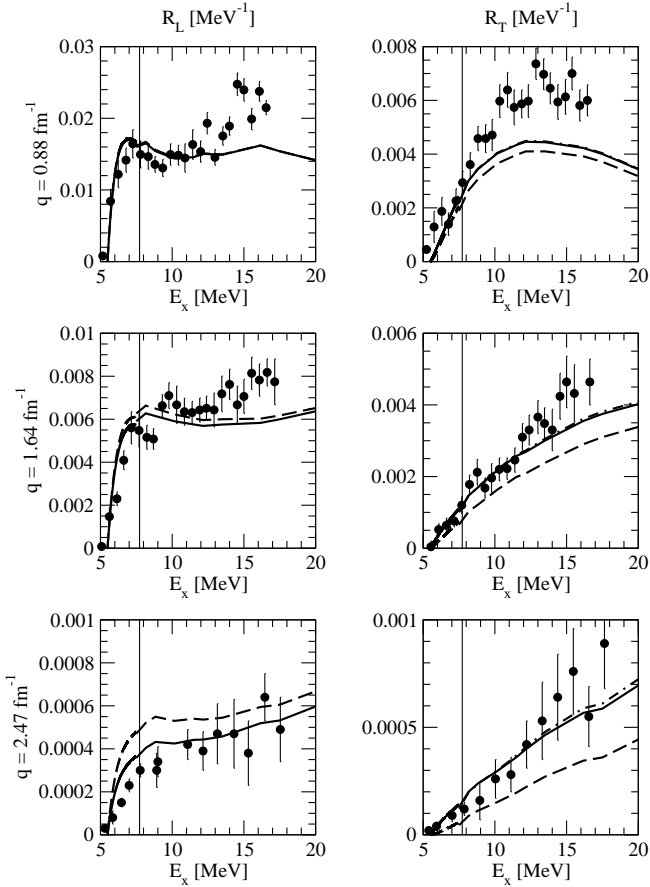}
}
\caption{Longitudinal and transverse response functions of $^3$He, 
obtained with the AV18/UIX Hamiltonian model.
The experimental data are from Ref.~\protect\cite{Ret94}. The vertical 
line represents the $ppn$ breakup threshold. See text for explanations of the 
different curves.}
\label{fig:rlrt}
\end{figure}

The theoretical predictions for the 
longitudinal and transverse response functions $R_L$ and $R_T$ for 
$q$=0.88, 1.64 and 2.47 fm$^{-1}$ and 5 MeV $\leq E_x\leq 20$ MeV 
are shown in Fig.~\ref{fig:rlrt}. The vertical lines 
indicate the DBT. Although the calculation is extended above this 
threshold, explicit three-body breakup channel contributions 
have not yet been included.
The dashed, dotted-dashed and 
solid lines are obtained with only one-body current and charge operators, 
with the one- and two-body operators of 
Ref.~\cite{Viv00}, and with the one- and many-body transition 
operators presented in Sec.~\ref{sec:em}. 
Note that in the case of the charge operator, the 
model of Ref.~\cite{Viv00} and the present one coincide. From inspection 
of the figure, we can observe that there is no significant difference 
between the results obtained using the model for the nuclear 
current operator of Ref.~\cite{Viv00} and the present one. 
To draw any conclusion in the comparison 
between theory and experiment, 
a complete calculation which includes also the $ppn$ channel
should be performed. 

\begin{figure}
\vspace*{0.4cm}
\resizebox{0.4\textwidth}{!}{%
\includegraphics{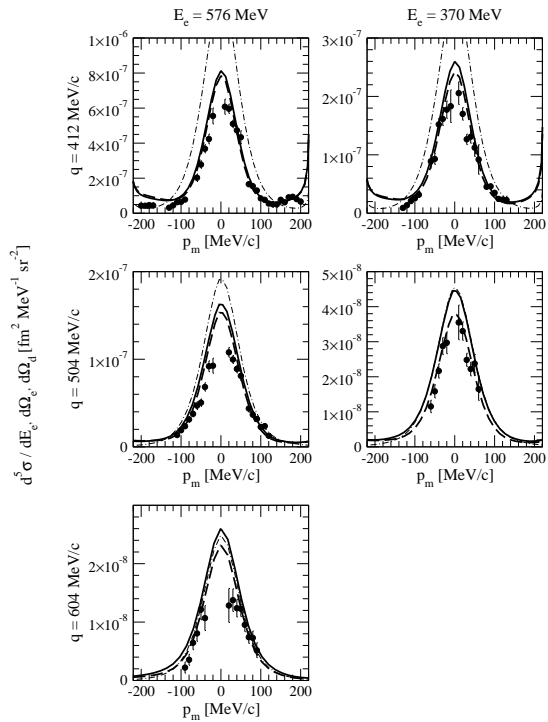}
}
\caption{Differential cross section of the 
$^3$He($e,e'd$)$p$ reaction  obtained with the AV18/UIX Hamiltonian model 
as function of the missing momentum at three 
$q$ values and two beam energies.
The data are from Ref.~\cite{Spa02}. See text for the explanation 
of the different curves.}
\label{fig:xsec}
\end{figure}

\begin{figure}
\vspace*{0.2cm}
\resizebox{0.48\textwidth}{!}{%
\includegraphics{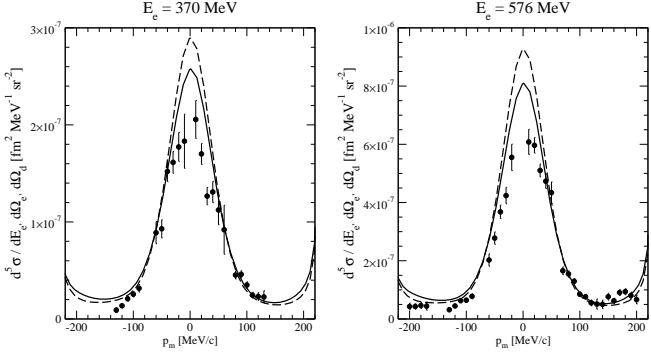}
}
\caption{Differential cross section of the 
$^3$He($e,e'd$)$p$ reaction  as function of the missing momentum at 
$q=$ 412 MeV/c and beam energy of 370 MeV and 576 MeV. 
The dashed and solid lines 
correspond to the calculation performed with the AV18 and with the 
AV18/UIX Hamiltonian models, respectively. 
Data are from Ref.~\cite{Spa02}.}
\label{fig:xsec_av18-av18uix}
\end{figure}

The theoretical predictions for the differential cross section of the 
$^3$He($e,e'd$)$p$ reaction  as function of the missing momentum 
at beam energies of 370 and 576 MeV 
and $q$ values of 412, 504 and 604 MeV/c are compared with 
the data of Ref.~\cite{Spa02} in Fig.~\ref{fig:xsec}. The dotted-dashed
lines correspond to the plane-wave impulse approximation results. 
When final-state-interaction effects are included and the $pd$ final 
wave function is calculated with the PHH technique using the AV18/UIX 
Hamiltonian model, the dashed and the solid lines are obtained, 
depending if the one-body only or the one- and many-body contributions 
to the nuclear transition operators are retained.
Here, no comparison is shown between 
the old model of Ref.~\cite{Viv00} and the present one, since no 
significant differences have been seen. Note that this is the 
first calculation for the $^3$He($e,e'd$)$p$ reaction above DBT 
which uses the PHH technique to calculate the bound- and scattering-state 
wave functions including TNI and Coulomb force effects in the 
final-state interaction.
By inspection of Fig.~\ref{fig:xsec}, 
we can conclude that there is an overall nice agreement between 
theory and experiment, although the data at low missing momentum are 
overestimated by the theory. This was already observed in Ref.~\cite{Spa02}, 
where the data were compared with a Faddeev calculation, with no 
Coulomb and three-nucleon interaction.
However, we have verified 
that the inclusion of the UIX three-nucleon 
interaction improves the description of the data at zero missing momentum.
This can be seen in Fig.~\ref{fig:xsec_av18-av18uix}, where the differential 
cross section at $q=$ 412 MeV/c and beam energy of 370 and 576 MeV is 
calculated with the AV18 (dashed curves) and the AV18/UIX 
(solid curves) Hamiltonian models. Only in this second case 
the model for the nuclear current operator includes 
three-body contributions, so that for each given Hamiltonian model the 
CCR is satisfied. 

\section{Summary and Outlook}
\label{sec:sum}

We have reported on new calculations of
$^3$He($e,e'$) longitudinal and transverse response functions 
for three values of the momentum transfer 
and excitation energies from two-body 
threshold up to 20 MeV, and for $^3$He($e,e'd$)$p$ 
differential cross section as function of the missing momentum in 
($q,\omega$)-constant kinematics, at two beam energies 
and three $q$ values. These calculations use accurate bound and 
scattering state wave functions obtained with the PHH method
from the Argonne $v_{18}$ two-nucleon and Urbana IX three-nucleon 
interactions. The model for the electromagnetic charge operator 
includes one- and two-body components, while the model for the electromagnetic 
current operator includes one-, two- and three-body 
components, constructed so as to 
satisfy {\it exactly} the CCR with the given 
Hamiltonian model. The model for the nuclear current operator has 
been tested calculating the $pd$ radiative capture cross section and spin 
observables. In particular, we have shown that the 
experimental $T_{20}$ and $T_{21}$ 
deuteron tensor observables are nicely reproduced by the 
present calculation. 
A systematic comparison between theory and experiment for 
the $pd$ radiative capture in a wide range of $E_{c.m.}$ 
is currently underway~\cite{Marip}.

The $^3$He electrodisintegration observables considered 
in the present work are 
very sensitive to the many-body contributions of the nuclear transition 
operators. However, no significant differences have been observed between 
the calculation performed with the present model of the nuclear 
current operator and the one of Ref.~\cite{Viv00}. 
In the case of the $^3$He($e,e'$) reaction, 
theoretical predictions and experimental data cannot be directly
compared above deuteron breakup threshold, since the 
(energetically open) three-body breakup channel contributions 
have yet to be calculated explicitly.
In the case of the $^3$He($e,e'd$)$p$ reaction, instead, 
the experimental differential cross section is fairly reproduced 
in the whole range of missing momentum for all different kinematics, when   
final-state interactions effects are considered. 
Further investigations for the 
$^3$He electrodisintegration in a wider energy range are vigorously 
being pursued.

The work of R.S. was supported by the U.S. DOE Contract 
No. DE-AC05-84ER40150, under which the Southeastern Universities 
Research Association (SURA) operates the Thomas Jefferson National 
Accelarator Facility.


\begin{thebibliography}{}
%
\bibitem{Car98} J.\ Carlson and R.\ Schiavilla, 
                Rev.\ Mod.\ Phys.\ \textbf{70}, 743 (1998).
%
\bibitem{Viv00} M.\ Viviani, A.\ Kievsky, L.E.\ Marcucci, S.\ Rosati, 
		and R.\ Schiavilla,
                Phys.\ Rev.\ C \textbf{61}, 064001 (2000).
%
\bibitem{Kieetal} A.\ Kievsky, M.\ Viviani, and S.\ Rosati,
                Nucl.\ Phys.\ {\bf A551}, 241 (1993); 
                Nucl.\ Phys.\ {\bf A577}, 511 (1994); 
		A.\ Kievsky, S.\ Rosati, and M.\ Viviani,
                Phys.\ Rev.\ C {\bf 64}, 024002 (2001).
%
\bibitem{Wir95} R.B.\ Wiringa, V.G.J.\ Stoks, and R.\ Schiavilla,
                Phys.\ Rev.\ C {\bf 51}, 38 (1995).
%
\bibitem{Pud95} B.S.\ Pudliner, V.R.\ Pandharipande, J.\ Carlson, 
		and R.B.\ Wiringa,
                Phys.\ Rev.\ Lett.\ {\bf 74}, 4396 (1995).
%
\bibitem{Ris85} D.O.\ Riska,
                Phys.\ Scr.\ {\bf 31}, 107 (1985);
		Phys.\ Scr.\ {\bf 31}, 471 (1985).
%
\bibitem{Sch89} R.\ Schiavilla, D.O.\ Riska, and V.R.\ Pandharipande,
                Phys.\ Rev.\ C {\bf 40}, 2294 (1989).
%
\bibitem{Mar03} L.E.\ Marcucci, M.\ Viviani, A.\ Kievsky, S.\ Rosati, 
                and R.\ Schiavilla, 
		Few-Body Syst.\ Suppl.\ {\bf 14}, 319 (2003);
		Few-Body Syst.\ Suppl.\ {\bf 15}, 87 (2003);
		M.\ Viviani, L.E.\ Marcucci, A.\ Kievsky, R.\ Schiavilla, 
		and S.\ Rosati, Eur.\ Phys.\ J.\ A {\bf 17}, 483 (2003).
%
\bibitem{Marip} L.E.\ Marcucci, M.\ Viviani, R.\ Schiavilla, 
                A.\ Kievsky, and S.\ Rosati, in preparation.
%
\bibitem{Sch90} R.\  Schiavilla, D.O.\  Riska, and V.R.\ Pandharipande, 
                Phys.\ Rev.\ C, {\bf 41}, 309 (1990).
%
\bibitem{Mar98} L.E.\  Marcucci, D.O.\ Riska, and R.\ Schiavilla, 
		Phys.\ Rev.\ C, {\bf 58}, 3069 (1998).
%
\bibitem{Ris89} D.O.\ Riska,
                Phys.\ Rep.\ {\bf 181}, 207 (1989).
%
\bibitem{Car90} J.\ Carlson, D.O.\ Riska, R.\ Schiavilla, and R.B.\ Wiringa,
                Phys.\ Rev.\ C {\bf 42}, 830 (1990).
%
\bibitem{Sac48} R.G.\ Sachs, Phys.\ Rev.\ {\bf 74}, 433 (1948);
                E.M.\ Nyman, Nucl.\ Phys.\ {\bf B1}, 535 (1967).
%
\bibitem{Coo79} S.A.\ Coon, {\it et al.},
                Nucl.\ Phys.\ {\bf A317}, 242 (1979).
%
\bibitem{Rob} M.R.\ Robilotta and M.P.\ Isidro Filho, 
              Nucl.\ Phys. {\bf A414}, 394 (1984); 
              Nucl.\ Phys. {\bf A451}, 581 (1986); 
	      M.R.\ Robilotta, M.P.\ Isidro Filho, H.T.\ Coelho, 
	      and T.K.\ Das, Phys.\ Rev.\ C {\bf 31}, 646.
%
\bibitem{Ret94} G.A.\ Retzlaff, {\it et al.}, 
                Phys.\ Rev.\ C {\bf 49}, 1263 (1994).
%
\bibitem{Spa02} C.M.\ Spaltro, {\it et al.},
                Nucl.\ Phys.\ {\bf A706}, 403 (2002).
%
\bibitem{Smi99} M.K.\ Smith and L.D.\ Knutson, 
                Phys.\ Rev.\ Lett.\ {\bf 82}, 4591 (1999).
%
\bibitem{Goe92} F.\ Goeckner, W.K.\ Pitts, and L.D.\ Knutson, 
                Phys.\ Rev.\ C {\bf 45}, R2536 (1992).
%
\end{thebibliography}
\end{document}